\documentclass[usenatbib,usegraphicx]{mn2e}
\include{epsf}

\title[{\tt GCD+}: A New Chemodynamical Approach to Modeling SNe and
  Chemical Enrichment]
{{\tt GCD+}: A New Chemodynamical Approach to Modeling Supernovae and
  Chemical Enrichment in Elliptical Galaxies }
\author[D.\ Kawata and B.K.\ Gibson]
 {D.~Kawata and B.K.~Gibson
\thanks{E-mail: dkawata,bgibson@astro.swin.edu.au}
\\
  Centre for Astrophysics and Supercomputing, 
  Swinburne University of Technology, Hawthorn VIC 3122, Australia
}
\date{Accepted .
      Received ;
      in original form }

\pagerange{\pageref{firstpage}--\pageref{lastpage}}
\pubyear{2002}

\begin{document}

\maketitle

\label{firstpage}

\begin{abstract}

 We have developed a new galactic chemo-dynamical evolution code,
called {\tt GCD+}, for studies of galaxy formation and evolution.
This code is based on our original three-dimensional tree
N-body/smoothed particle hydrodynamics code which
includes self-gravity, hydrodynamics, radiative cooling,
star formation, supernova feedback, and metal enrichment.
{\tt GCD+} includes a new Type II (SNe II) and Ia (SNe Ia)
supernovae model taking into account the lifetime of progenitor stars, and
chemical enrichment from intermediate mass stars.
We apply {\tt GCD+} to simulations of elliptical galaxy formation, and
examine the colour-magnitude relation (CMR), the Kormendy relation,
and the [Mg/Fe]--magnitude relation of simulation end-products.
{\tt GCD+} is a useful and unique tool which enables us 
to compare simulation results with
the observational data directly and quantitatively.
Our simulation confirm the results of Kawata (2001)
who uses a simpler chemo-dynamical evolution code.
We newly find that radiative cooling becomes
more efficient and thus the gas infall rate increases, with decreasing
mass of galaxies, which contributes to the slope of the CMR.
In addition, the sophisticated treatments of both SNe II and 
SNe Ia in {\tt GCD+} show that feedback from SNe Ia plays a crucial role
in the evolution of elliptical galaxies. We conclude that
the feedback effect of SNe Ia should not be ignored
in studying the evolution of elliptical galaxies.

\end{abstract}


\section{Introduction}
\label{sintro}

 The abundances of elements heavier than helium (``metals'')
in interstellar gas and in stars vary systematically from place to
place within galaxies. Since heavy elements are believed to have been
synthesized in stars, the metal abundances in present-day
galaxies should offer a record through which we may trace the
history of star formation within galaxies.
Since the pioneering work of \citet{bt72}, studies of
chemical evolution have succeeded in connecting observed metal abundances
to the history of star formation within galaxies
\citep[e.g.][]{ay87, mt87, tww95, bg97, cmg97, tcbmp98}.
Moreover, recently developed semi-analytic models \citep{wr87, wf91}
including chemical evolution make it possible to relate 
the observed chemical properties of galaxies to cosmological theory
\citep[e.g.][]{kc98, clbf00, on01, spf01, bbfsf02}.
These techniques require low computational costs
and allow large coverage of parameter space.
Although the pure chemical evolution models and the
semi-analytic models seem to succeed in
explaining the observational properties of galaxies at various redshifts,
they inevitably assume a phenomenological model involving
a number of parameters to describe the processes of gas accretion rates,
radiative cooling, star formation, and supernova feedback 
within a galactic halo.
In addition, structures of the end-products can not be discussed,
because there is no information about the dynamical history.

On the other hand, recent advances in computer technology and numerical methods
have made it possible to calculate the dynamical and chemical evolution
of galaxies self-consistently \citep[e.g.][]{clc98,mtlc01,cc02}.
Three dimensional chemo-dynamical evolution codes have been
applied to the following studies over the past five years:
\citet{sm95} studied the chemo-dynamical evolution of
disk galaxies and succeeded in distinguishing the chemical properties
between halo, bulge, and disk stars 
\citep[see also][]{bc00,bc01}.
\citet{rvn96} and \citet{pb99} took into account 
metal enrichment by Type Ia supernovae (SNe Ia)
as well as Type II supernovae (SNe II), 
and reproduced the correlation 
between [O/Fe] and [Fe/H] for stars in the solar neighbourhood.
Such self-consistent calculation of chemical evolution
allows one to analyse the photometric properties of simulation end-products
in combination with stellar population synthesis, and
to obtain the luminosity of the end-products
without the assumption of an arbitrary mass to light ratio. 
Taking advantage of such technique, \citet{sn99}, \citet{ns00},
and \citet{ksw00} discussed the zero-point of the Tully-Fisher relation
of disk galaxies. \citet{bs98,bs99,bs00,bs01} 
studied the dynamical and photometric
properties of elliptical galaxies formed by a merger of two disk galaxies.
\citet{mytn97} showed that the galactic wind caused by supernovae (SNe)
feedback leads to a positive colour gradient observed in some
dwarf elliptical galaxies. Using cosmological simulation, 
\citet{ng98}, \citet{co99}, \citet{tlmc01}, and \citet{tvk02}
studied metal enrichment of the intergalactic medium
\citep[see also][as a work coupling semi-analytic model to
hydro simulations]{ahs01}, and \citet{nfco01} examined
the luminosity function and colour distribution function
of galaxies at different redshifts.
Since radiative cooling depends on the metallicity,
dynamical evolution is quite sensitive to
the chemical evolution \citep{kh98,kpj00}. Thus, we must 
calculate chemical and dynamical evolution self-consistently.

 Using a chemo-dynamical evolution code, \citet[][hereafter K01b]{dk01b}
were able to reproduce the colour-magnitude relation (CMR) of 
observed elliptical galaxies. K01b showed that 
SNe feedback affects the evolution of lower mass systems more strongly,
and induces a stronger galactic wind in lower mass systems.
Such low mass systems have lower metallicities and bluer colours than
higher mass systems, as predicted by analytic models 
\citep[e.g.][]{ay87,bg97}
and more phenomenological numerical simulations
\citep[e.g.][]{rl74,rc84}.
Moreover, K01b found that galactic winds are
triggered mainly by SNe Ia rather than SNe II, although 
K01b ignored the lifetime of progenitors of SNe II and SNe Ia
(i.e.\ K01b assumes the instantaneous recycling approximation
for SNe II, and all SNe Ia occur simultaneously 1.5 Gyr after stars were born).

We have developed a new chemo-dynamical evolution code, called {\tt GCD+},
which adopts a more sophisticated chemical evolution model than that 
discussed in K01b. 
We have relaxed the instantaneous recycling approximation for SNe II,
and adopt a more sophisticated SNe Ia model 
\citep[][hereafter KTN00]{ktn00}.
Our code takes into account the metallicity dependent lifetime of stars
\citep{tk97}, and metal enrichment from intermediate mass stars 
\citep{vdhg97}. 
The purpose of this paper is to introduce details of {\tt GCD+}.
In addition, as an application of {\tt GCD+}, we carry out simulations
of elliptical galaxy formation similar to the ones of K01b. 
Then we re-examine the CMR, the Kormendy relation,
and the [Mg/Fe]--magnitude relation, which were studied in K01b.
Since these observed relations are well-established,
and are expected to reflect the chemo-dynamical evolution 
of elliptical galaxies, this is an important application for {\tt GCD+}.
Next, taking advantage of the sophisticated SNe II and SNe Ia
models in {\tt GCD+}, we clarify the role of SNe Ia, compared to
SNe II, in the evolution of elliptical galaxies.

The outline of this paper is as follows.
Details of {\tt GCD+} are described in Section \ref{scode}.
In Section \ref{sapegf}, we use {\tt GCD+} to simulate
elliptical galaxy formation.
We show the galaxy formation model used in this paper
briefly in Section \ref{smodel}.
In Sections \ref{scmr}--\ref{smgfe},
results of numerical simulations are presented,
and are compared to the observed scaling relations.
Discussion about the study of elliptical galaxy formation
is given in Section \ref{sdisc}.
Finally, we present our conclusions from this paper in Section \ref{sconc}.

\begin{figure}
  \leavevmode
  \epsfxsize=80mm  
  \epsfbox{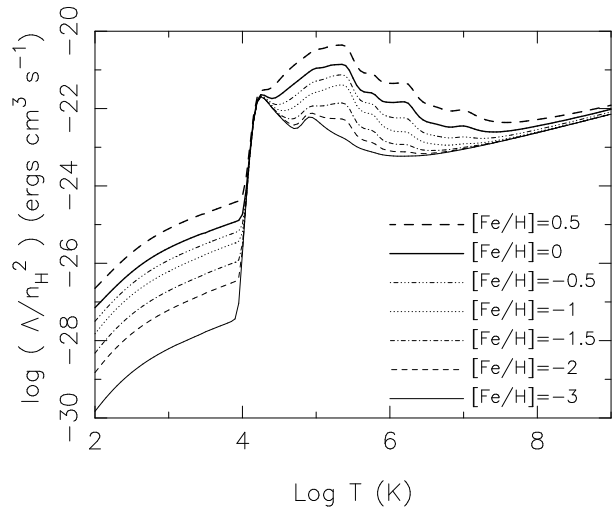}
 \caption{ Cooling rates as a function of temperature.
 Each curve corresponds to a cooling rate with different metallicities
 as indicated. }
 \label{cr-fig}
\end{figure}

\section{The New Chemo-Dynamical Evolution Code}
\label{scode}

 The new code, {\tt GCD+} is an update version of the code used in \citet{dk99}
and K01b. The code is essentially based on TreeSPH \citep{hk89,kwh96},
which combines the tree algorithm \citep{bh86}
for the computation of the gravitational forces with the smoothed
particle hydrodynamics \citep[SPH:][]{l77,gm77}
approach to numerical hydrodynamics.
The dynamics of the dark matter
and stars is calculated by the N-body scheme, and the
gas component is modeled using the SPH.
It is fully Lagrangian, three-dimensional, and highly adaptive in space
and time owing to individual smoothing lengths and
individual time steps. Moreover, it includes self-consistently
almost all the important physical processes
in galaxy formation, such as self-gravity, hydrodynamics,
radiative cooling, star formation, SNe feedback, and
metal enrichment. The code is vectorized and parallelized.
The parallelization is done using the MPI library,
so that the code can run on any type of computer, including
supercomputers and PC clusters.
We have revised the modeling of cooling, star formation,
and SNe feedback drastically from the code used in K01b. 
Here, we will only deal with the new aspects of our N-body/SPH
code, which has already been described in \citet{dk99}.

\subsection{Gas Cooling}

 To model gas cooling, we use the cooling curves computed by 
MAPPINGS III\footnote{MAPPINGS III is the successor of MAPPINGS II
described in \citet{sd93}, and available at 
http://www.mso.anu.edu.au/$^\sim$ralph/map.html}
written by R.S. Sutherland.
Using MAPPINGS III, we have computed the cooling function of the
ionization 
equilibrium gas with various metallicities.
Fig.~\ref{cr-fig} shows all the cooling curves which we use.
With linear interpolation of these cooling curves, 
the code calculates the cooling rate of $\Lambda(T,{\rm [Fe/H]})$ as a function
of the temperature and metallicity for each gas particle.
We assume that the cooling rate of the gas with metallicity
of lower than ${\rm [Fe/H]}=-3$ is the same as the one with ${\rm [Fe/H]}=-3$,
and the cooling rate of the gas with ${\rm [Fe/H]}>0.5$ is
equivalent to that at ${\rm [Fe/H]}=0.5$ to avoid extrapolation.
The cooling rate for the gas with the solar metallicity
is larger than that for the gas of primordial composition
by more than an order of magnitude, thus
 cooling by metals should not be ignored
in numerical simulations of galaxy formation
\citep{kh98,kpj00}.
The temperature for each particle is derived by 
$T_i= P_i \mu m_p/(k_{\rm B} \rho_i)$, where $P_i$ and $\rho_i$, 
are the pressure and density for $i$-th particle, and $\mu$,
$k_{\rm B}$, and $m_p$ are the mean molecular weight, Boltzmann's
constant, and the proton mass, respectively. 
For simplicity, we fix $\mu=0.6$, regardless of the metallicity.
We set the lower limit of the temperature to be $T_{\rm lim}=100$ K.

\subsection{Star Formation and Initial Mass Function}
\label{ssf}

 We model star formation using a method similar to
that suggested by \citet{nk92} and \citet{kwh96}.
The criteria for star formation and the star formation rate
are the same as those in K01b.
We use the following three criteria for star formation:
1) the gas density is greater than a critical density,
$\rho_{\rm crit} = 2 \times 10^{-25}\ {\rm g\ cm^{-3}}$,
i.e.\ $n_{\rm H} \sim 0.1 {\rm cm^{-3}}$; 
2) the gas velocity field is convergent,
${\bf \nabla} \cdot \mbox{\boldmath $v$}_i < 0$; and 3) the 
Jeans unstable condition, $h/c_s>t_{\rm g}$, is satisfied,
here $h$, $c_s$, and $t_{\rm g} = \sqrt{3 \pi/16 G \rho_{\rm g}}$ 
are the SPH smoothing length,  
the sound speed, and the dynamical time of the gas respectively
\citep{dk99}.
 When a gas particle is eligible to form stars,
its star-formation rate (SFR) is
\begin{equation}
 \frac{d \rho_*}{dt} = -\frac{d \rho_{\rm g}}{dt}
 = \frac{c_* \rho_{\rm g}}{t_{\rm g}},
\label{sfreq}
\end{equation}
where $c_*$ is a dimensionless SFR parameter
and $t_{\rm g}$ is the dynamical time, 
which is longer than the cooling timescale
in the region eligible to form stars.
This formula corresponds to the Schmidt law that
SFR is proportional to $\rho_{\rm g}^{1.5}$.
We fix $c_*=0.5$, following K01b.

 We assume that stars, which are represented by a star particle,
are distributed according to the \citet{es55} initial mass function (IMF).
The IMF by number, $\Phi(m)$, in each mass interval $dm$
is defined as\footnote{Equation (\ref{imf-eq}) corresponds
to an IMF by mass $\propto m^{-x}$}
\begin{equation}
 \Phi(m) dm = Am^{-(1+x)} dm,
\label{imf-eq}  
\end{equation}  
where $x=1.35$ is the Salpeter index and the coefficient $A$
is determined by the normalization in the mass range
$M_{\rm l} \leq m \leq M_{\rm u}$.
We set $M_{\rm l}=0.2\ {\rm M_{\rm \sun}}$
and $M_{\rm u}=60\ {\rm M_{\rm \sun}}$ in this paper.

\begin{figure}
  \leavevmode
  \epsfxsize=80mm  
  \epsfbox{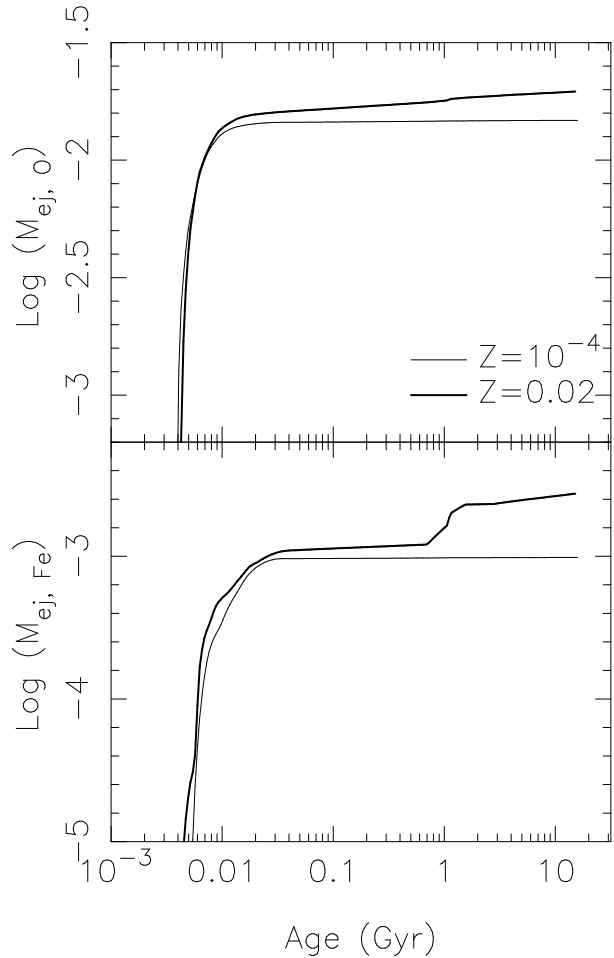}
 \caption{ Chemical yields as a function of age of a star particle
with mass of 1 M$_{\rm \sun}$.
The upper (lower) panel shows the total ejected oxygen (iron) mass.
The thick (thin) line indicates the history of 
a star particle with the metallicity of $Z=0.02$ $(10^{-4})$.
 }
 \label{cy-fig}
\end{figure}

\subsection{Feedback}
\label{sfd}

 {\tt GCD+} takes account of energy feedback and metal enrichment 
by SNe, and metal enrichment from intermediate mass stars. 
We consider here both SNe II and SNe Ia. 
The code calculates the event rates of SNe II and SNe Ia,
and the yields of SNe II, SNe Ia, and intermediate mass stars
for each star particle at every time step,
considering the IMF and stellar lifetimes.
Here, we assume the same stellar lifetimes as the ones used in 
\citet{tk97} and \citet{ka97}.
The simulation follows the evolution of the abundances
of several chemical elements ($^1$H, $^4$He,$^{12}$C, $^{14}$N, $^{16}$O, 
$^{20}$Ne, $^{24}$Mg, $^{28}$Si, $^{56}$Fe, and Z,
where Z means the total metallicity). We refer to Table 2 of \citet{ww95}
for the solar abundances.

\subsubsection{Type II Supernovae}
\label{ssneii}

 We assume that each massive star ($\geq8\ {\rm M_{\rm \sun}}$)
explodes as a Type II supernova.
To calculate the ejected mass of gas and metals by SNe II,
we use the stellar yields derived by \citet{ww95}.
\citet{ww95} provide yields for a grid of stellar masses
between 11 and 40 M$_{\rm \sun}$, and metallicities between
0 and 1 Z$_{\rm \sun}$. With linear interpolation
of these grids, we obtain yields as a function of
the stellar mass and metallicity. \citet{ww95} considered three different
models for $m\geq30$ M$_{\rm \sun}$, their A, B, and C models.
These differ in the amount of energy imparted by the piston in their
models at explosion initiation. Following \citet{tww95}, we use
the 'B' model in this mass regime. For $m\geq40$ M$_{\rm \sun}$, 
we assume the same abundance ratios for the yields and the same mass fraction
for the remnants as those for 40 M$_{\rm \sun}$ star.
For $Z\geq Z_{\rm \sun}$, we use the following simple scaling,
e.g.\ the yield of carbon from a star with the mass, $m$, and metallicity, $Z$,
\begin{eqnarray}
 m_{ej,C}(Z,m) & = & m_{ej,C}(Z_{\rm \sun},m)+
  m_{ej} (Z_{\rm \sun},m) Z_{C,{\rm \sun}} \nonumber \\
& &\times (Z/Z_{\rm \sun}-1),
\label{mejc-eq}
\end{eqnarray}
where $m_{ej,C}$ is the total mass of ejected carbon, including newly
synthesized and initial carbon;
$m_{ej}=m_{ej,H}+m_{ej,He}+m_{ej,Z}$ 
indicates the total ejected mass; $Z_{C,{\rm \sun}}$ is the 
solar abundance of carbon.
The yields from stars with initial masses between 8 and 11 M$_{\rm \sun}$
are still unclear \citep[e.g.][]{hin93}.
Hence, we simply interpolate linearly between the yields for
the lowest mass, i.e.\ $\sim$ 11 M$_{\rm \sun}$, star in \citet{ww95}
and those for the highest mass, i.e.\ 8 M$_{\rm \sun}$ star in
\citet{vdhg97}, as mentioned in the next section.

\subsubsection{Intermediate Mass Stars}
\label{simms}

 The yields and remnant masses for intermediate mass stars ($\leq8$
M$_{\rm \sun}$) are taken from \citet{vdhg97}. \citet{vdhg97} present
the yields of $^1$H, $^4$He, $^{12}$C, $^{13}$C, $^{14}$N, and $^{16}$O
for stars with initial masses between 0.8 and 8 M$_{\rm \sun}$
and initial metallicities of $Z=0.001-0.04$. We do not use $^{13}$C
yields. \citet{vdhg97} provide the newly formed and ejected metals
unlike \citet{ww95} who present the total ejected metals.
Therefore, we calculate the total ejected metals
using Table 1 of \citet{vdhg97} for H, He, C, N, and O and
the initial abundances for the other elements, i.e.\
Ne, Mg, Si, and Fe, which are simply scaled to the solar abundance set.
As for stars with $Z>0.04$, we use the same
scaling as equation (\ref{mejc-eq}) based on $Z=0.04$ yields.
For $Z<0.001$, we simply assume the same yields as $Z=0.001$ yields.
Whilst this simple assumption may overestimate the ejected metals,
since the metals are ejected from intermediate mass stars later
than those from SNe II, which eject much larger amounts of metals,
this simplification should not affect the final metallicity. 
Nevertheless, the knowledge of the yields for low and zero metallicity
stars is important in studies of the chemical composition
of extremely metal poor objects observed at both low and high redshifts,
although it is not yet well-established \citep[e.g.][]{fii00,cdls01,mgcw01}.

\subsubsection{Type Ia Supernovae}
\label{ssneia}

 We adopt the SNe Ia model proposed by KTN00,
which is based on the SNe Ia progenitor scenario 
proposed by \citet{hkn99}. This model is different from
the conventional one proposed by \citet{gr83}.
Details of this model are described in KTN00.
Here, we briefly explain this model and implementation in our code.
Following KTN00, SNe Ia are assumed to occur at binary systems which consist of
primary and companion stars with appropriate masses and metallicities
(${\rm [Fe/H]}\geq-1.1$). The primary stars have
the main-sequence mass in the range of $m_{p,l}=3$ M$_{\rm \sun}$
and $m_{p,u}=8$ M$_{\rm \sun}$, and evolve into the C+O white dwarf (WD).
The mass range of companion stars are restricted to 
between $m_{d,RG,l}=0.9$ M$_{\rm \sun}$ and $m_{d,RG,u}=1.5$ M$_{\rm \sun}$ 
and between
$m_{d,MS,l}=1.8$ M$_{\rm \sun}$ and $m_{d,MS,u}=2.6$ M$_{\rm \sun}$.
KTN00 designate the binary system whose companion has the mass within the 
former (latter) mass range ``RG+WD (MS+WD) system''.
Finally, the total number of SNe Ia is obtained by the following equation
as a function of age, $t$, of a star particle 
with the mass of $m_s$ M$_{\rm \sun}$,
\begin{eqnarray}
 N_{\rm SNe Ia}(t) & = & m_s \int_{m_{p,l}}^{m_{p,u}} m^{-(1+x)} dm
 \left\{ \int_{M_l}^{M_u} m^{-x} dm \right\}^{-1} \nonumber \\ 
 & & \times  \left[
  b_{\rm MS} \frac{\int_{\max(m_{d,MS,l},m_t)}^{m_{d,MS,u}}
  \Phi_d (m) dm }
  { \int_{m_{d,MS,l}}^{m_{d,MS,u}}
  \Phi_d (m) dm } \right. \nonumber \\
 & & + \left.
  b_{\rm RG} \frac{\int_{\max(m_{d,RG,l},m_t)}^{m_{d,RG,u}}
  \Phi_d (m) dm }
  { \int_{m_{d,RG,l}}^{m_{d,RG,u}} \Phi_d (m) dm }\right],
\label{nia-eq}
\end{eqnarray}
where $m_t$ is the mass of star whose lifetime is equal to $t$,
and the mass function of the companion stars 
is assumed to be $\Phi_d (m) \propto m^{-1.35}$ by number,
following KTN00. 
The term before the square bracket indicates the number of 
C+O WDs, i.e.\ primary stars. 
The first term within the square bracket determines 
how much fraction of C+O WDs evolves into SNe Ia from
the MS+WD systems, and the second term 
determines how much fraction of C+O WDs evolves into SNe Ia from
the RG+WD systems. 
Following KTN00, we set $b_{\rm MS}=0.05$
and $b_{\rm RG}=0.02$.
The nucleosynthesis prescriptions for SNe Ia are taken
from the W7 model of \citet{ibn99}.
  
 Fig.~\ref{cy-fig} shows the total amount of oxygen and iron ejected from
a star particle with the mass of 1 M$_{\rm \sun}$ as a function of its age.
Initially, metals are ejected only by SNe II. This continues until
the 8 M$_{\rm \sun}$ star dies ($\sim 0.04$ Gyr in the case of $Z=0.02$).
Oxygen is produced mainly by SNe II. 
After SNe II cease, the continuous ejection of oxygen is mainly due to 
the contribution from intermediate mass stars.
On the other hand, a significant fraction of iron is produced by SNe Ia.
Stars with $Z=0.02$ lead to SNe Ia depending on the lifetime of
their companion stars. SNe Ia occur in MS+WD (RG+WD) systems 
in the range of age of $0.7-1.5$ ($> 2.8$) Gyr.
However, a star particle with $Z=10^{-4}$ does not lead to SNe Ia,
because the above SNe Ia model restricts the metallicity range for
progenitors of SNe Ia to ${\rm [Fe/H]}\geq-1.1$. 
For simplicity, we assume that the metallicity range for SNe Ia 
is $\log Z/Z_{\rm \sun}\geq-1.1$, instead of ${\rm [Fe/H]}\geq-1.1$.
In the implementation, the code possesses a look-up table
of the yields of all the chemical elements, remnant masses, number of SNe
as a function of the age and metallicity, and 
calculates those values for each star particle at every time step.

\subsubsection{Energy Feedback}
\label{sef}

 One of the most difficult and most critical processes to model
in galaxy formation simulations is the way in which the energy feedback from
SNe affects the surrounding gas. Unfortunately, there is no
clear understanding of how this should be modeled.
We adopt a simple model proposed by \citet{nw93} and used in K01b.
This model assumes that the energy produced by SNe affects only
the temperature and velocity field of the surrounding gas,
and its effect is implemented by increasing 
the thermal ($E_{\rm th}$) and kinetic ($E_{\rm kin}$)
energy of the gas neighbours of each star particle
by an amount corresponding to the energy released by SNe.
We assume that a parameter $f_v=E_{\rm kin}/(E_{\rm kin}+E_{\rm th})$ 
defines the fraction of the available energy 
to perturb the gas velocity field, and the rest of the energy 
of the SNe contributes to the increase in the thermal energy 
of the gas (see K01b for details).
It is known that kinetic feedback affects the history of 
star formation more strongly 
than thermal feedback, which quickly dissipates
due to efficient radiative cooling where the gas density is
high enough to form stars.
The parameter, $f_v$, controls the magnitude of the effect
of SNe \citep[K01b;][]{nw93}.
We assume that each SN yields the energy of $\epsilon_{SN} \times
10^{51}$ ergs, and then
$E_{\rm kin}=f_v \epsilon_{SN} 10^{51}$ ergs. 
Because an initial SN energy has not been established quantitatively yet,
we consider the available SN energy to be a free parameter.
Finally, in our code, there are two parameters, $f_v$ and
$\epsilon_{SN}$, to control the magnitude of the effect of SNe. 

\subsubsection{Implementation}
\label{impf}

 Based on the above feedback processes, the code calculates
the amount of the mass, energy, and heavy elements 
released from each star particle within each time step.
Our code adopts the simple feedback scheme suggested by \citet{nk92}
\citep[see also][for a proposed alternative scheme]{lpc02}.
The mass, energy, and heavy elements are smoothed over the neighbouring
gas particles using the SPH smoothing algorithm.
For example, when the $i$-th star particle ejects the mass of
$M_{\rm SN,{\it i}}$, the increment of the mass of the $j$-th neighbour 
gas particle is given by
\begin{equation}
 \Delta { M_{\rm SN,{\it j}}}
  =  \frac{m_j}{\rho_{{\rm g},i}} { M_{\rm SN,{\it i}}}
  W(r_{ij}/h_{i}),
\label{msneq}
\end{equation}
where
\begin{equation}
 \rho_{{\rm g},i} = \langle \rho_{\rm g}(\mbox{\boldmath $x$}_i) \rangle
 = \sum_{j \neq i} m_j W(r_{ij}/h_{i}),
\end{equation}
and $W(x)$ is an SPH kernel, 
\begin{equation}
W(x) = \frac{8}{\pi h^{3}}
 \left\{ \begin{array}{cc}
 1-6x^{2}+6x^{3} & {\rm if}\ 0\leq x\leq 1/2, \\
 2(1-x)^{3}      & {\rm if}\ 1/2\leq x\leq 1, \\
 0               & {\rm otherwise}.
\end{array} \right.
\label{Weq}
\end{equation}
Here, $r_{ij}$ is the distance between the $i$-th and $j$-th particles,
and $h_{i}$ is the smoothing length for the $i$-th particle.
Note that the above equations employ $h_{i}$, 
instead of $h_{ij}=(h_{i}+h_{j})/2$
which is adopted in other SPH calculations, such as density and
pressure gradient \citep[see][]{dk99}. This is because feedback is
a one-way process, which does not have to be symmetrised.
In addition, if all neighbour particles have much smaller smoothing 
length than the $i$-th particle, $h_{ij}$ is possible to become smaller 
than $r_{ij}$, and then the $i$-th particle loses the place to feedback.
The use of $h_i$ ensures to avoid this problem.

We revise the update algorithm for 
the smoothing length in \citet{dk99} who uses the algorithm suggested by
\citet{hk89}. The new algorithm is based on that suggested by \citet{ttpct00}.
First, we count neighbour particles for the $i$-th particle with $h_i$
using the following smoothed kernel,
\begin{equation}
W_{nn}(r/h_i) =
 \left\{ \begin{array}{cc}
 1  & {\rm if}\ 0\leq r/h_i\leq 3/4, \\
 \frac{\pi h_i^3}{8} W(4(r/h_i-3/4)) & {\rm if}\ 3/4\leq r/h_i\leq 1,
\end{array} \right.
\label{wnneq}
\end{equation}
where $r$ is the distance of the neighbour particle from the $i$-th particle,
and $W(x)$ is the same spline kernel as equation (\ref{Weq})
\footnote{Note that our smoothing definition is different from that
used in \citet{ttpct00}, in other words our smoothing length is twice as
large as theirs.}. The number of neighbours, $N_{{\rm nb},i}$, is 
no longer integer, but real number. 
This revision helps alleviate the discontinuity 
in the number of neighbours.
 
 Next, to avoid a sudden change in the smoothing length,
we modify equation (14) of \citet{dk99}
which determines the evolution
of the smoothing length. Setting $s=(N_s/N_{{\rm nb},i}^n)^{1/3}$,
equation (14) of \citet{dk99} is expressed as,
\begin{equation}
 h_i^{n+1}=h_i^n (1-a+a s),
\label{heq}
\end{equation}
where $N_s=40$ is the desired number of neighbours,
$h_i^n$ means the smoothing length of the $i$-th particle at $n$ step.
Equation (14) of \citet{dk99} corresponds to the case of $a=0.5$.
Here we change $a$ as a function of $s$,
\begin{equation}
a = 
 \left\{ \begin{array}{cc}
 0.2 (1+s^2) & {\rm if}\ s<1, \\
 0.2(1+1/s^3) & {\rm if}\ s\geq 1.
\end{array} \right.
\label{aeq}
\end{equation}

 In addition, \citet{ttpct00} use a modified Courant condition
depending on the relative velocities to neighbour particles.
Because this condition requires a significant amount 
of additional computational cost,
once star formation and kinetic feedback are involved, we abandon
this condition.
As shown in Fig.~1 of \citet{ttpct00}, the above two 
algorithms lead to a dramatical improvement in keeping the number 
of neighbour particles constant, and the modified Courant
condition adds a marginal effect.

We follow the evolution of the smoothing length of star particles
until the end of simulation, because we relax the instantaneous
recycling for SNe II and consider the feedback and chemical enrichment
from SNe Ia and intermediate mass stars.
Even in the above new algorithm, involving kinetic feedback 
enables gas particles to escape from the position of
star particles more quickly than the speed of the increase 
in the smoothing length of the star particles. 
In serious cases, it happens that there is no neighbour particle, 
and the mass conservation is broken due to the missing
of the mass deposition from stars to gas. To solve this problem,
we introduce the iteration process using equation (\ref{heq}), 
once the number of neighbour particles is less than ten. 
Although this iteration requires an extra computational
cost, it ensures complete mass conservation.

\section{ Application to Elliptical Galaxy Formation}
\label{sapegf}

 To see how the new code works, in this section, 
we apply {\tt GCD+} to simulations 
of elliptical galaxy formation in the semi-cosmological model
described in K01b.
K01b examined the mass-dependent properties of elliptical galaxies, 
such as the CMR, the Kormendy relation, and the [Mg/Fe]--magnitude
relation.
Our new code relaxes some of the simple assumptions in K01b,
and becomes a more sophisticated code.
Hence, the re-examination of those properties is an interesting application 
as well as an important test for {\tt GCD+}.

\begin{figure}
  \leavevmode
  \epsfxsize=80mm  
  \epsfbox{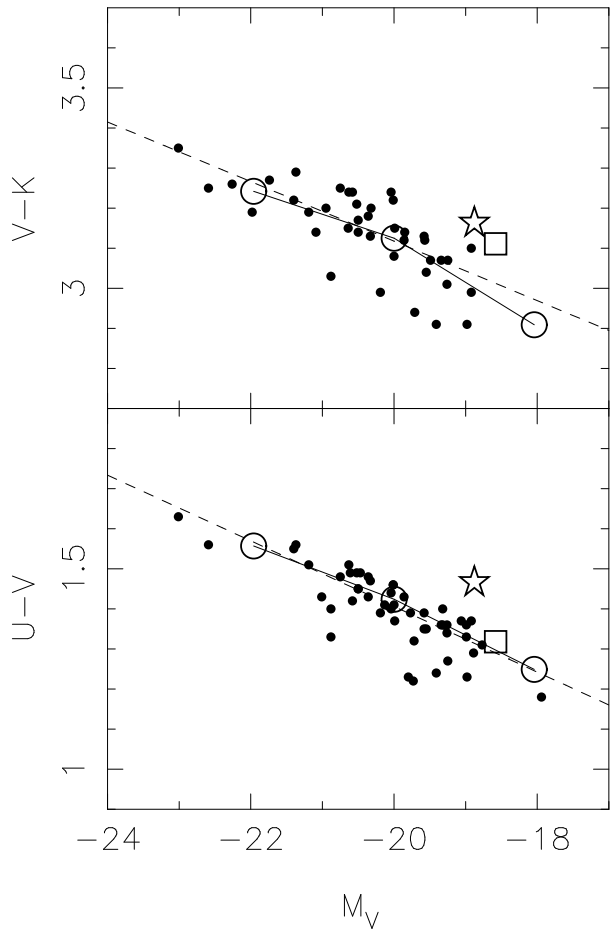}
 \caption{ Comparison of the CMRs for the simulation end-products
 and Coma Cluster galaxies \citep[small dots,][]{ble92a}
 in the aperture of 5 kpc. The circles connected by
 solid lines indicate the CMRs for the best model.
 The square (star) denotes the result of model noIa-LM (iIIbIa-HM)
 in Section \ref{seia} (\ref{seiiia}).
 The dashed line shows the CMR fitted to the Coma Cluster galaxies 
\citep{ble92b}.
 }
  \label{cmr-fig}
\end{figure}

\begin{figure}
  \leavevmode
  \epsfxsize=80mm  
  \epsfbox{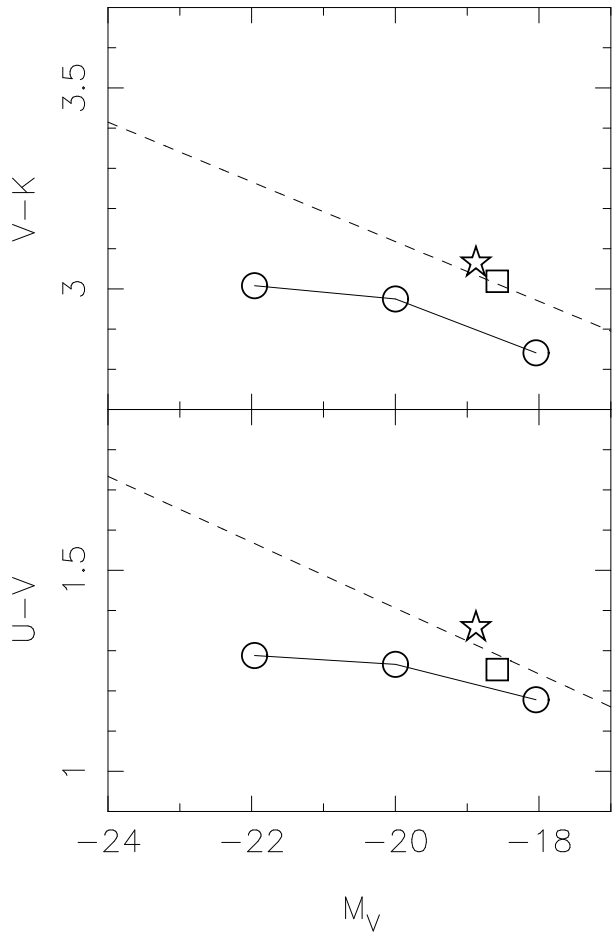}
 \caption{ 
 The CMRs for the simulation end-products
 in the 99 kpc aperture. The circles connected by
 solid lines indicate the results of the best model.
 The square (star) denotes the result of model noIa-LM (iIIbIa-HM).
 The dashed lines are the same as those in Fig.~\ref{cmr-fig}.
 }
  \label{lacmr-fig}
\end{figure}

\subsection{Elliptical Galaxy Formation Model}
\label{smodel}

Following K01b, we consider an isolated sphere, as a seed galaxy,
on which small-scale density fluctuations
corresponding to a cold dark matter (CDM) power spectrum are superimposed.
Here, we use Bertschinger's software COSMICS \citep{eb95}
to generate initial density fluctuations.
To incorporate the effects of fluctuations with longer wavelengths,
the density of the sphere has been enhanced and a rigid rotation
corresponding to a spin parameter, $\lambda$, has been added.
The initial conditions of this model are determined
by the following four parameters: $\lambda$, $M_{\rm tot}$,
$\sigma_{\rm 8,in}$, and $z_{\rm c}$.
The spin parameter is defined by
\begin{equation}
 \lambda \equiv \frac{J|E|^{1/2}}{G M_{\rm tot}^{5/2}},
\end{equation}
where $J$ is the total angular momentum of the system,
$E$ is the total energy,
and $M_{\rm tot}$ is the total mass of the sphere, which is
composed of dark matter and gas;
$\sigma_{\rm 8,in}$ is the rms mass fluctuation in a sphere of radius
$8\ h^{-1}$ Mpc, which normalizes the amplitude of the CDM power spectrum;
$z_{\rm c}$ is the expected collapse redshift.
If the top-hat density perturbation has an amplitude of
$\delta_i$ at the initial redshift, $z_i$, we obtain
$z_{\rm c} = 0.36 \delta_i (1+z_i)-1$ approximately
\citep[e.g.][]{tp93}.
Thus, when $z_{\rm c}$ is given, $\delta_i$ at $z_i$ is determined.
\citet{dk99} found that
the seed galaxy which has a slow rotation corresponding to $\lambda=0.02$
and small-scale density fluctuations evolves into an elliptical-like
system. Thus, we employ $\lambda=0.02$.
This spin parameter is close to the minimum value possible
in a CDM universe, according to the results of N-body simulations
\citep{be87,wqsz92}. 
We fix $\sigma_{\rm 8,in}=0.5$ and $z_c=3.5$.
Our simulations assume a flat universe ($\Omega = 1$)
with a baryon fraction of $\Omega_{\rm b} = 0.1$
and a Hubble constant of $H_0 = 50\ {\rm km\ s^{-1}\ Mpc^{-1}}$.
We carry out each simulation using 9171 particles for gas and dark matter,
respectively. 
We simulate the evolution of each model from $z_i=40$ to $z=0$.

 The morphological evolution of all the models which are shown
in this paper are similar to the evolution seen in Fig.~1 of \citet{dk99}. 
A nearly spherical stellar system 
is formed at $z=0$ in all the models.
Following K01b, we focus on the chemical and photometric properties
for the simulation end-products at $z=0$.
In our simulations the stellar particles each carry their own age and
metallicity ``tag'', which, when combined with population synthesis,
enables us to derive the photometric properties of the simulated stellar
systems. The photometric properties are derived by the same analysis
as K01b. Details of this analysis are described in 
Section 3.2 of \citet{dk01a} and Section 4 of K01b.

In this analysis, the spectral energy distribution (SED) of each
stellar particle is assumed to be that of a single stellar population (SSP)
that means a coeval and chemically homogeneous assembly of stars.
Since the observational data with which our results should be compared
provide the luminosity distribution projected to a plane,
we have to derive the projected distribution of SED
from the three dimensional distribution of stellar particles.
Finally, we obtain the $x$-$y$ projected images as shown in Figure 5 of 
\citet{dk01a}, when $z$-axis is set to be the initial rotation axis. 
We confirmed that the results do not depend on the direction of the
projection, because the end-products are nearly spherical.
The flux of each stellar particle is
smoothed using a Gaussian filter with the filter scale of 1/4 of the
softening length of the stellar particle.
These images provide similar information to the imaging data obtained in
actual observations. Thus, we can obtain various photometric properties from
these images in the same way as in the analysis of observational imaging data.
We use the images similar to the one displayed in Figure 5 of \citet{dk01a},
but employing a 1001 $\times$ 1001 pixel mesh
to span the squared region with 100 kpc on a side.

We adopt the data of SSPs of Kodama \& Arimoto 97 model \citep{tk97,ka97}.
Kodama \& Arimoto 97 model supplies the database of SSPs with two types of IMF:
($x$, $M_{\rm u}$, $M_{\rm l}$) = (1.35, 60, 0.1) and
($x$, $M_{\rm u}$, $M_{\rm l}$) = (1.1, 60, 0.1)
in the definition of Section \ref{ssf}.
We adopt the data of SSPs with
the IMF of ($x$, $M_{\rm u}$, $M_{\rm l}$) = (1.35, 60, 0.1),
while we use the IMF of ($x$, $M_{\rm u}$, $M_{\rm l}$)
= (1.35, 60, 0.2) in the numerical simulations.
As mentioned in Section 4 of K01b, this inconsistency
does not affect the final photometric properties.

 The global photometric properties, such as the total luminosity,
the colours, and the effective radius, are obtained from the
projected image data.
Following K01b, the total magnitude and the effective radius
are derived by fitting the surface brightness profile 
to the $r^{1/n}$ law (eq.\ [11] of K01b).
We set the center to the position of a pixel
which has the maximum $V$ band luminosity.

These procedures enable us to compare the properties
for simulation end-products with those for observed elliptical
galaxies directly and quantitatively.
 In the following sections, we begin by showing the best model
which can reproduce the CMR of observed elliptical galaxies, and
examine what is responsible for the CMR based on the detailed analysis
of the simulation results.
Next, we discuss the physical sizes of the simulation end-products,
comparing them with the observed Kormendy relation.
Finally, we examine the correlation between the total magnitude and
the abundance ratio of Mg to Fe, i.e.\ the [Mg/Fe]--magnitude relation,
which is sensitive to the star formation history.

\begin{figure}
  \leavevmode
  \epsfxsize=80mm  
  \epsfbox{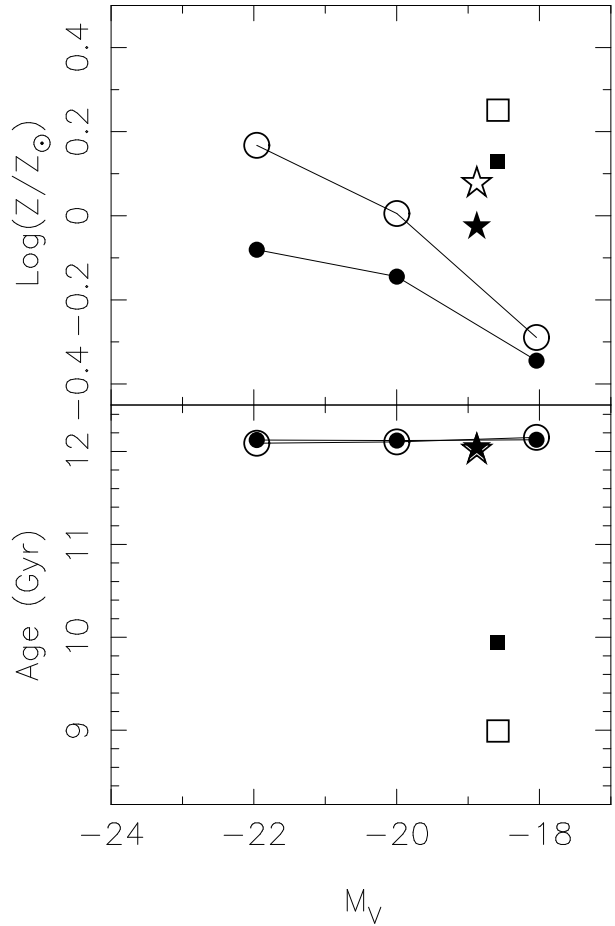}
 \caption{ 
  The metallicities (upper panel) and ages (lower panel)
 against the absolute $V$ band magnitude for each model.
 The circles connected by solid lines indicate the results of the best model.
 The square (star) denotes the result of model noIa-HM (iIIbIa-LM).
 The open (filled) symbols denote the values evaluated in the 5 kpc
 (99 kpc, spread over almost the whole galaxy) aperture.
 }
  \label{zage-fig}
\end{figure}

\begin{figure*}
  \leavevmode
  \epsfxsize=160mm  
  \epsfbox{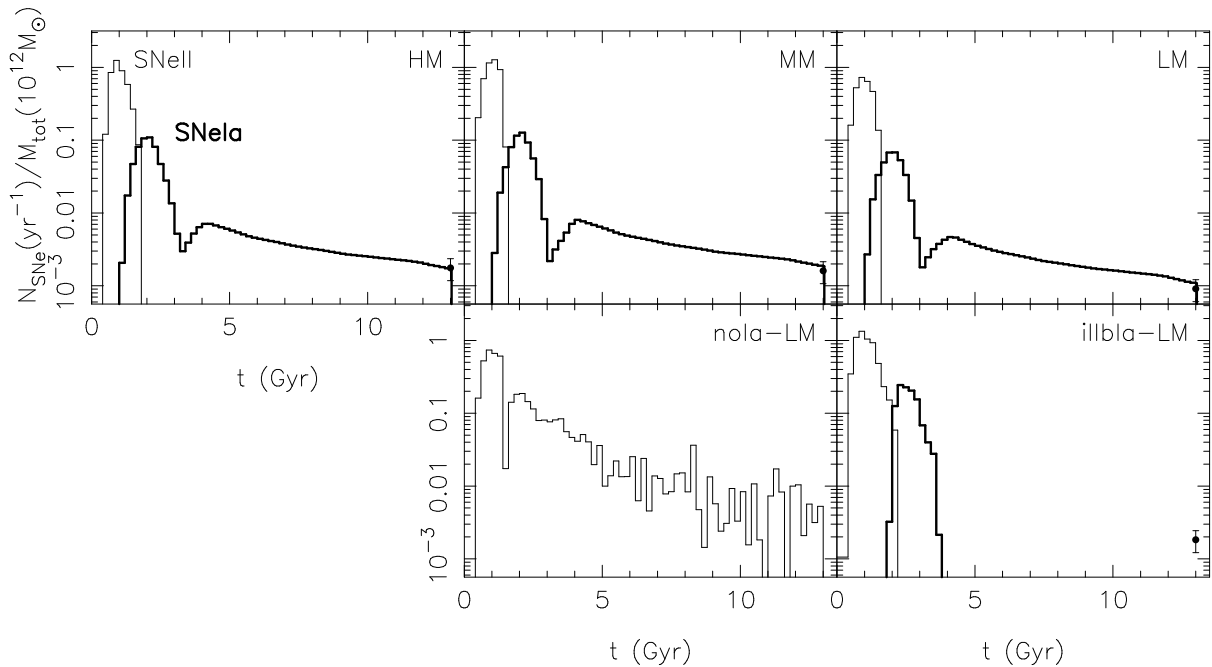}
 \caption{ 
  Time variation of the event rate of
 SNe II (thin lines) and SNe Ia (thick lines) for all the models.
 Points with error bars (at $t=13$ Gyr) are taken from the observational 
 SNe Ia rate by \citet{cet99}. 
 }
  \label{snhr-fig}
\end{figure*}

\begin{table*}
 \centering
 \begin{minipage}{140mm}
 \caption{Model Parameters}
 \label{mod-tab}
 \begin{tabular}{@{}lccccccc}
 & & \multicolumn{2}{c}{Particle Mass (M$_{\rm \sun}$)} & 
 \multicolumn{2}{c}{Softening (kpc)} &  \\
 Name & $M_{\rm tot}$ (M$_{\rm \sun}$) & Dark Matter &
 Gas & Dark Matter & Gas & SNe II & SNe Ia \\
 HM & $4\times10^{12}$ & $3.93\times10^8$ & $4.36\times10^7$ &
 4.28 & 2.06 & yes & yes \\
 MM & $8\times10^{11}$ & $7.85\times10^7$ & $8.72\times10^6$ &
 2.50 & 1.20 & yes & yes \\
 LM & $2\times10^{11}$ & $1.96\times10^7$ & $2.18\times10^6$ &
 1.58 & 0.758 & yes & yes \\
 noIa-LM & $2\times10^{11}$ & $1.96\times10^7$ & $2.18\times10^6$ &
 1.58 & 0.758 & yes & no \\
 iIIbIa-LM & $2\times10^{11}$ & $1.96\times10^7$ & $2.18\times10^6$ &
 1.58 & 0.758 & instantaneous & 1.5 Gyr delay \\
 \end{tabular}
 \end{minipage}
\end{table*}

\begin{table*}
 \centering
 \begin{minipage}{140mm}
 \caption{Global Photometric Properties: Surface Brightness Profiles and
 Gradients}
 \label{gpp-tab}
 \begin{tabular}{@{}lcccccccc}
 & \multicolumn{3}{c}{$V$ band} & 
 \multicolumn{3}{c}{$K$ band} &  \\
  & $M_V$ & & $r_e$ & $M_K$ & & $r_e$ & $[\Delta (B-R)]$
 & $[\Delta
  \log(Z/Z_{\rm \sun})]$ \\
 Name & (mag)& $n$ & (kpc) & (mag) & $n$ & (kpc) & $/\Delta \log r$
  & $/\Delta \log r$ \\
 HM &  $-21.96$ & 5.80 & 15.0 & $-$24.94 & 6.51 & 11.9 &
  $-$0.17 &  $-0.38$ \\
 MM &  $-20.00$ & 3.63 & 5.69 & $-$22.94 & 3.86 & 4.84 &
  $-$0.14 &  $-0.30$ \\
 LM &  $-18.04$ & 1.80 & 3.33 & $-$20.88 & 1.71 & 3.08 &
  0.00 &  $-$0.01 \\
 noIa-LM & $-$18.58 & 3.01 & 1.32 & $-$21.62 & 2.86 & 1.09 & 0.10 & $-$0.79 \\
 iIIbIa-LM & $-$18.88 & 7.26 & 2.34 & $-$21.90 & 7.47 & 1.64 & 
  $-$0.12 & $-$0.32 \\
 \end{tabular}
 \end{minipage}
\end{table*}

\begin{table*}
 \centering
 \begin{minipage}{140mm}
 \caption{Photometric Properties and Stellar populations
within Apertures}
 \label{ppsp-tab}
 \begin{tabular}{@{}lcccccccccc}
 & \multicolumn{4}{c}{$D<5$ kpc} & 
 \multicolumn{4}{c}{$D<99$ kpc} & \multicolumn{2}{c}{$D<r_{e,B}/2$ } \\
  & & & & Age & & & & Age & \\
 Name & $U-V$ & $V-K$ & [Z/H] & (Gyr) & 
 $U-V$ & $V-K$ & [Z/H] & (Gyr) & [Mg/Fe] & [E/Fe] \\
 HM & 1.56 & 3.24 & 0.17 & 12.1 & 1.29 & 3.01 & $-$0.08 & 12.1 &
  0.06 & 0.10 \\
 MM & 1.42 & 3.13 & 0.01 & 12.1 & 1.27 & 2.97 & $-$0.14 & 12.1 & 
  0.10 & 0.13 \\
 LM & 1.25 & 2.91 & $-$0.29 & 12.2 & 1.18 & 2.84 & $-0.34$ & 12.1 & 
  0.14 & 0.15 \\
 noIa-LM & 1.32 & 3.11 & 0.25 & 8.99 & 1.25 & 3.02 & 0.13 & 9.94 & 
  0.05 & 0.04\\
 iIIbIa-LM & 1.47 & 3.16 & 0.08 & 12.0 & 1.36 & 3.07 & $-0.03$ & 12.0 &
  0.08 & 0.11
 \end{tabular}
 \end{minipage}
\end{table*}

\subsection{The Best Model: the Colour-Magnitude Relation}
\label{scmr}

 As explained in Section \ref{scode}, due to the lack of knowledge
of the physics of star formation and SNe feedback,
our code has a number of free parameters,
such as the mass range of the IMF ($M_{\rm l}$ and $M_{\rm u}$ in
Section \ref{ssf}) and the feedback parameters of $f_v$ and
$\epsilon_{SN}$ in Section \ref{sef}. 
We fix these parameters so that the observed CMR is reproduced.
To compare with the CMR, we simulate the evolution of 
seed galaxies with three different masses of $M_{\rm tot}=4\times 10^{12}$,
$8\times 10^{11}$, and $2\times 10^{11}$ 
M$_{\rm \sun}$ (hereafter, models HM, MM, and LM, respectively). 
The mass and spatial resolutions of each model are summarized in
Table \ref{mod-tab}.

The mass range of the IMF controls yields per mass of new born star
particles.
Higher $M_{\rm u}$ or $M_{\rm l}$ leads to larger yields,
and makes the metallicity of the end-products higher.
We chose $(M_{\rm u}, M_{\rm l})=(60, 0.2)$ M$_{\rm \sun}$, which
produces enough yields for the highest mass model (HM)
to reproduce the observed red colour of galaxies with similar luminosities.
As shown in K01b, we also find that the 
stronger feedback causes the steeper slope in the CMR.
The magnitude of the feedback effect is controlled by the parameters
$\epsilon_{SN}$ and $f_v$.
We fix the parameter set of $(\epsilon_{SN}, f_v)=
(0.1, 0.002)$ such that the observed CMR is reproduced, when 
the above IMF is adopted.
According to high resolution 1D simulations of an SN remnant
in \citet{tgjs98}, 90 \% of the initial SN energy is lost in radiation
during its early expansion phase; this is not resolved in our simulations.
The available SN energy of $10^{50}$ ergs which we adopt
is consistent with the result of \citet{tgjs98},
when a canonical initial SN energy of $10^{51}$ ergs is assumed.
\citet{tgjs98} also suggested that in the last stage of their simulation
about $\sim 8.5 \times10^{49}$ erg, i.e.\ 85 \% of the available energy,
is found in the form of kinetic energy. This corresponds to $f_v=0.85$
in our definition, and is much larger than that we assumed here.
However, the resolution limit of our simulation (a few kpc) is 
much larger than the physical scale of their simulation ($\sim$0.15 kpc).
Unfortunately, little is know about the fraction of the kinetic
energy which is still available to affect the interstellar medium 
on such large scales. 
Therefore, $f_v$ suggested by \citet{tgjs98} is considered to 
be an upper limit, which does not invalidate the lower value 
which we adopted here.
Although this parameter set is not a unique solution, 
it is one which best explains the observed CMR.
Therefore, the set of models HM, MM, and LM with the above parameters 
is called the ``best model'' in this paper.

\begin{figure}
  \leavevmode
  \epsfxsize=80mm  
  \epsfbox{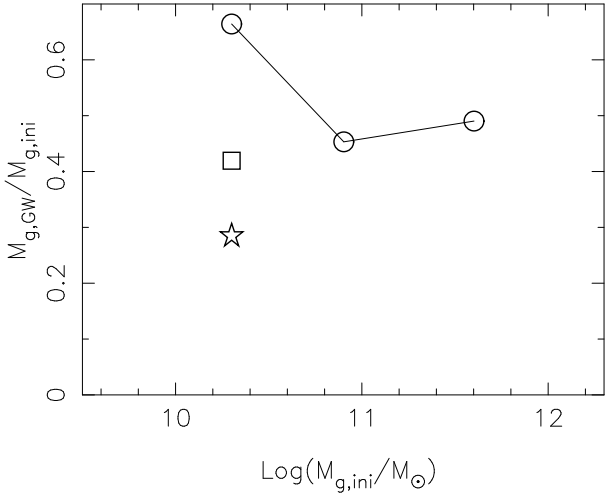}
 \caption{ 
 The ratio of the ejected gas mass ($M_{\rm g,GW}$)
 at $z=0$ to the initial gas mass ($M_{\rm g,ini}$).
 The circles connected by solid lines indicate the results of the best model.
 The square (star) denotes the result of model noIa-LM (iIIbIa-LM).
 Here, the ejected gas is defined as the sum of all the gas particles whose
 galactocentric radius is greater than 20 kpc.
 }
  \label{gwm-fig}
\end{figure}

 Fig.~\ref{cmr-fig} shows the CMR of the best model which reproduces
the observed CMR for galaxies in the Coma Cluster very well.
The data for galaxies in the Coma Cluster are the observed CMR of 
\citet{ble92a}.
Since there is no difference between S0s and ellipticals
in the scaling relations which we now discuss,
we do not distinguish between S0s and ellipticals.
\citet{ble92a} supply $U-V$ and $V-K$ colours
which refer to an aperture size of 11 arcsec, and
the $V$ band total magnitude derived from a combination of
their data and the literature.
Throughout this paper we adopt the distance modulus of the Coma
cluster of $m-M=34.7$ mag;
the Virgo distance modulus of $m-M=31.01$ \citep{gffkm99}
and the relative distance modulus of the Coma with respect to the Virgo
of $m-M=3.69$ \citep{ble92b}.
This gives the luminosity distance of 87.1 Mpc for the Coma.
We assume that the angular diameter distance equals
the luminosity distance, because the redshift of the Coma
cluster ($z\sim0.023$) is nearly zero cosmologically.
Then the aperture size of 11 arcsec at the distance of the Coma Cluster
corresponds to $\sim$ 5 kpc. 
The colours for simulation end-products are derived
in the same aperture size as that used in the observations.

K01b claimed that the aperture effect should not be ignored,
when discussing the CMR observed in an aperture of a small size.
Observed elliptical galaxies (especially in luminous galaxies)
have colour gradients which make the colour at the centre redder than that in
the outer regions \citep[e.g.][]{pdidc90}.
The end-products in our simulations for modes HM and MM
also have significant colour gradients as seen in Table \ref{gpp-tab}.
Since colours within a fixed aperture yield
colours in a more central region for larger galaxies,
it is possible that the colours of large galaxies become 
redder than those of small galaxies,
even if the {\it mean} colour of the whole galaxy is the 
same between the large and small galaxies.
Fig.~\ref{lacmr-fig} shows the CMR for the mean colour
within the aperture of 99 kpc, which covers almost the entire galaxy
in all the models which we have examined.
As expected, the larger galaxy's colours are changed more dramatically 
by the aperture, which makes the slope between models HM and
MM shallower than that between models MM and LM. 
The aperture effect contributes significantly
to the slope of the CMR in Fig.~\ref{cmr-fig}
in the mass range between models HM and MM \citep[see also][]{kaba98}.

 As seen in K01b, Table \ref{gpp-tab} shows that the colour and
metallicity gradients become shallower, and Sersic law index $n$ 
(see eq.\ [11] of K01b) in 
the surface brightness profile becomes smaller, with decreasing mass
of the system. It may appear strange that the mean colour within the
99 kpc aperture is bluer than that within the 5 kpc aperture
in model LM, although the colour gradient for model LM is flat.
This is because Table \ref{gpp-tab} shows the gradients 
within the radius where the $B$-band magnitude 
is brighter than $\mu_B=24.5$ mag arcsec$^{-2}$, following K01b, 
and the gradient at larger radii becomes negative.

 The sequences of colours and line strengths among
elliptical galaxies can almost equally well be attributed to either age
difference or metallicity differences \citep{gw94}.
In our simulation this degeneracy can be broken completely,
because we can directly analyse the metallicity and age in the simulation
end-products. Fig.~\ref{zage-fig} shows the metallicity and age
for each model. Here, the metallicities and ages are luminosity weighted
values. Although the ages are all very old irrespective of the models and
the apertures, the metallicities exhibit similar behavior to 
the colours as a function of luminosity.
Thus, we conclude that the slope of the CMR of the best model
in Fig.~\ref{cmr-fig} is caused solely by the changes in metallicity.

\begin{figure*}
  \leavevmode
  \epsfxsize=160mm  
  \epsfbox{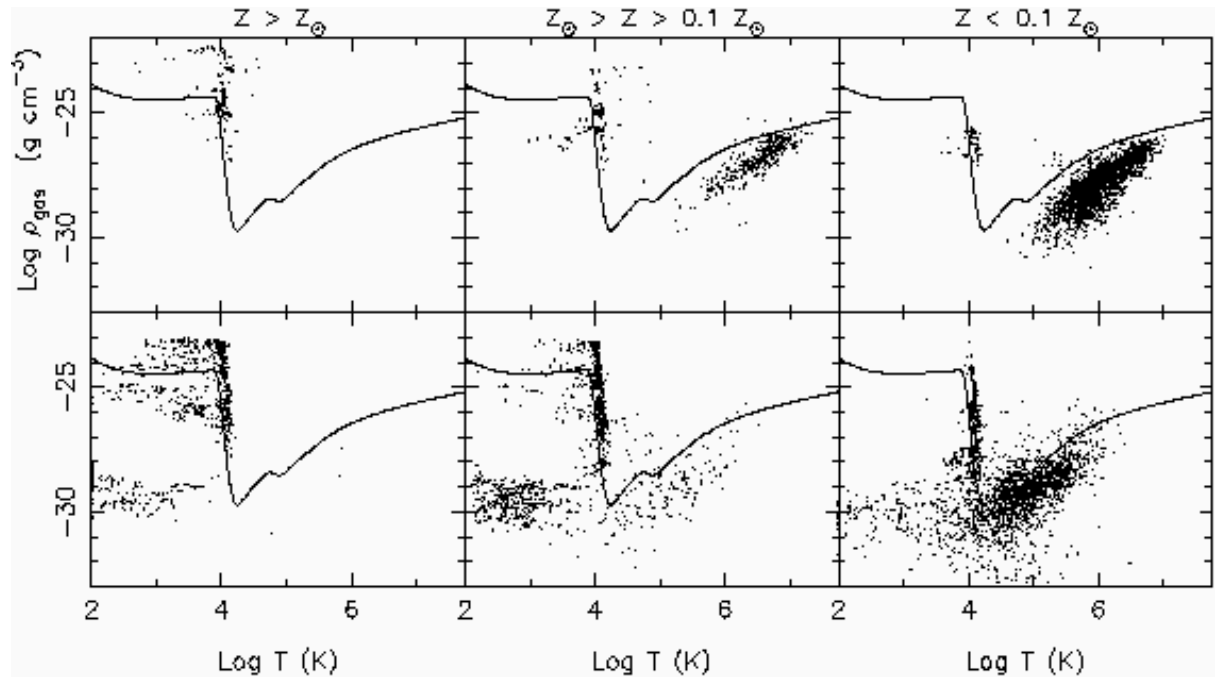}
 \caption{ 
  The density-temperature distribution of the gas particles 
 with higher (left), intermediate (middle), and lower (right) metallicity 
 at $z=3.27$ for models HM (upper panels) and LM (lower panels). 
 The solid curves separate the region where the cooling times are 
 shorter (upper region) and longer than the age of the universe at $z=3.27$.
 }
  \label{tcdyn-fig}
\end{figure*}

Fig.~\ref{snhr-fig}
shows the histories of SNe II and SNe Ia event rates normalized 
by the total mass of the system.
In our simulations, the histories of SNe II events roughly trace
those of star formation, because the lifetime of SNe II progenitors
is shorter than the size of the bins of the histograms.
We can clearly see that
star formation ceases abruptly around $t=1.7$ Gyr ($z=1.7$),
regardless of the mass of the system.
We confirmed that gas particles begin to blow out from
the stellar system, i.e.\ the galactic wind occurs
when star formation stops.
Subsequently, all the gas particles overcome
the binding energy of the dark matter and stars, and escape from the
system, as shown in Fig.~9 of K01b. 

\begin{figure}
  \leavevmode
  \epsfxsize=80mm  
  \epsfbox{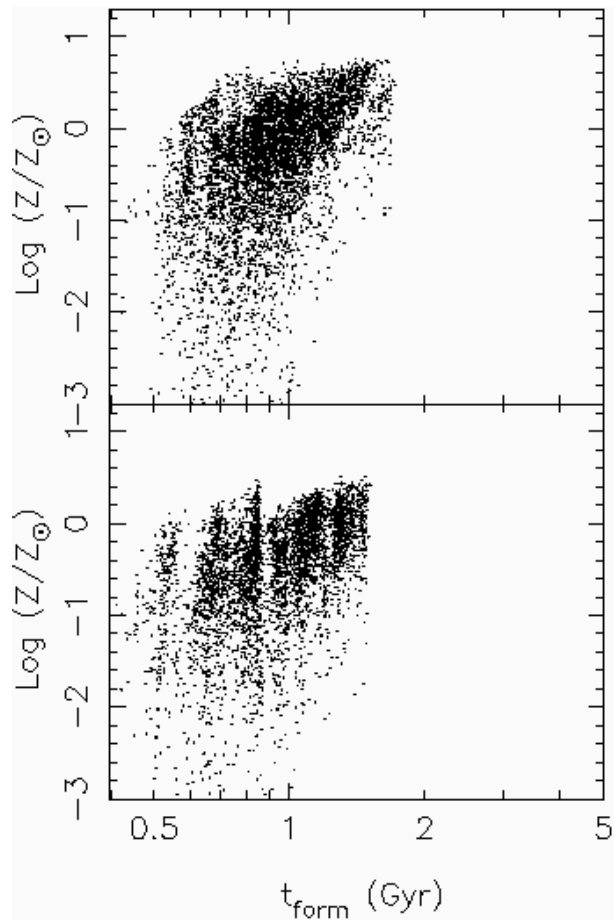}
 \caption{ Metallicities against age for the star particles 
 in the final stellar systems of models HM (upper panel) 
 and LM (lower panel). }
  \label{meta-fig}
\end{figure}

Fig.~\ref{gwm-fig} shows the mass fraction of the ejected gas 
as a function of the mass of the system. 
The ejected gas mass in Fig.~\ref{gwm-fig} is defined as the total mass of
all the gas particles whose galactocentric radius is greater than 
20 kpc at $z=0$. Comparing between models MM and LM, the lower mass system
ejects a greater fraction of gas.
Since the ejected gas cannot contribute to further metal enrichment,
metal enrichment is more strongly suppressed in the lower mass system,
leading to its lower metallicity in Fig.~\ref{zage-fig}. 
On the other hand, between models HM and MM,
the lower mass system ejects a slightly smaller fraction
of the gas. In this luminosity range, the aperture effect appears to be
a dominant factor in explaining the slope of the mass-metallicity
relation. However, there is a significant slope between models HM and MM
in mean metallicity within the large aperture (Fig.~\ref{zage-fig}). 
We find that this is because there is another factor which contributes to 
the mass-metallicity relation.
 Fig.~\ref{tcdyn-fig}
plots density against temperature for the gas particles
at $z=3.27$ for models HM and LM. At $z=3.27$ the system is almost
relaxed, the gas particles which stay in the lower right region
in the panel are hot gas which has not yet cooled since they 
virialised. On the other hand, the gas particles which stay in the upper 
left ($\log \rho_{\rm gas}\geq-28$ and $\log T\leq4$) region
represent the cold gas. The gas particles in the lower left region are
escaping from the system. 
Comparing between models, the high mass system has more hot gas
and less cold gas than the low mass system. The solid lines in the panels 
indicate where the cooling time equals the age of the universe at $z=3.27$.
Cooling is efficient in the upper region from this line, and inefficient
in the lower region from this line. In the high mass system,
the hot gas particles fall into where cooling is inefficient.
Therefore, a large fraction of the gas cannot cool or infall into the central 
region where stars are forming, thus the hot gas is not
affected by metal enrichment from stars (Fig.\ \ref{tcdyn-fig}).
Since the amount of gas is small in the central region, 
relative to the amount of metals returned from stars,
the metallicity of the gas in the central region increases rapidly.
In other words, a low infall rate of the gas induced by 
inefficient cooling leads to efficient chemical enrichment.
This is a similar mechanism to the infall model
which was proposed to solve the G-dwarf problem
of the closed box chemical evolution model for the solar vicinity
\citep[e.g.][]{bejp97}.
On the other hand, in the low mass system the hot gas 
stays close to the line in Fig.~\ref{tcdyn-fig}.
A large fraction of this gas which cools efficiently
infalls into the central region, and suffers from metal enrichment.
Since there is a large amount of gas in the star forming
region, the metallicity of the gas in the central region slowly increases. 
In other words, a high infall rate of gas, caused by efficient cooling,
leads to inefficient chemical enrichment. Consequently, the high mass system
makes metal rich stars more quickly than the low mass system
(Fig.~\ref{meta-fig}), and the mean metallicity of the high mass system
becomes higher than that of the low mass system.
Hence, we conclude that not only the galactic wind, which ejects
a larger fraction of gas in the lower mass system,
but also the mass dependence of the gas infall rate,
contributes to the slope in
the mass-metallicity relation (Fig.\ \ref{zage-fig}) and 
the CMR (Fig.\ \ref{cmr-fig}).

\subsection{The Role of SNe Ia}
\label{seia}

To assess the relative importance of SNe Ia to SNe II during
the evolution of elliptical galaxies,
we have run an additional model
which is similar to model LM, except that SNe Ia are not included.
Hereafter, this model is called ``noIa-LM''.
Figs.~\ref{cmr-fig}--\ref{gwm-fig} also show results of model noIa-LM.
Comparison between models LM and noIa-LM
demonstrates that SNe Ia are crucial sources
for the galactic wind. For example, Fig.~\ref{snhr-fig} 
shows that model noIa-LM
does not have a clear cessation of star formation.
Consequently, model noIa-LM ejects a
smaller fraction of gas from the system (Fig.~\ref{gwm-fig}), 
and has a higher metallicity and a redder colour than model LM
(Figs.~\ref{cmr-fig} and \ref{zage-fig}). 
In the $U-V$ and $M_V$ diagram, model noIa-LM appears to follow the CMR
of the best model. This is due to the younger age of model noIa-LM
as seen in the lower panel of Fig.~\ref{zage-fig}.
Model noIa-LM adopts the same feedback parameters as those of model LM.
In addition, model noIa-LM has a larger number of SNe than model LM,
because more stars have formed. Thus, this result indicates that
the galactic wind is caused mainly by SNe Ia. 
It is also notable that the event rate of SNe Ia exceeds that of SNe II,
when the galactic wind occurs in models HM, MM, and LM. We therefore 
conclude that SNe Ia play a crucial role in driving 
and maintaining the galactic wind,
and thus in contributing to the evolution of elliptical galaxies.
This conclusion is consistent with K01b's suggestion, and
our sophisticated treatment of SNe II and SNe Ia 
reinforces this idea.

\subsection{Instantaneous recycling SNe II and
 burst SNe Ia model}
\label{seiiia}

 K01b adopted instantaneous recycling for SNe II and
a simple burst SNe Ia model, in which all the SNe Ia occur
simultaneously 1.5 Gyr after star formation.
To compare the feedback effect of this simple description of SNe in K01b
with that of the more realistic SNe model used in this paper,
we now examine another model which employs
instantaneous recycling SNe II and 1.5 Gyr delay burst SNe Ia.
The total number of SNe II is given by the integral of number
of stars with the mass higher than 8 M$_{\rm \sun}$.
The total number of SNe Ia expected in each star particle 
is assumed to be given by the sum of the MS+WD and RG+WD
binary systems, which corresponds to the integral of equation (\ref{nia-eq})
from $t=0$ to $t=\infty$. 
We also assume that
SNe Ia occur only in star particles with $\log Z/Z_{\rm \sun}\geq-1.1$,
following the KTN00 model.
Using this SNe model, we carry out the simulation with the same total mass
as model LM. Hereafter, this model is called ``iIIbIa-LM''.

Fig.~\ref{snhr-fig} shows that the galactic wind ceases
star formation in this model, although the amount of ejected gas
is much smaller than that for model LM (Fig.~\ref{gwm-fig}).
As a result, model iIIbIa-LM has a higher metallicity
and a redder colour than model LM (Figs.~\ref{cmr-fig} and \ref{zage-fig}).
In both models LM and iIIbIa-LM, the galactic wind occurs just after
SNe Ia ignite. However, SNe Ia  in model iIIbIa-LM begin to occur 
later than those in model LM, due to the simple 1.5 Gyr delay,
and model iIIbIa-LM has a longer duration of star formation
than model LM. As a result, in model iIIbIa-LM the chemical enrichment
progresses for a longer time, which leads to the higher metallicity
and redder colour.
Thus, the final properties of elliptical galaxies
are sensitive to the time delay of SNe Ia, which is the main trigger
of the galactic wind. It is also worth noting that
the peak rate of star formation in model iIIbIa-LM
is significantly higher than that in model LM (Fig.~\ref{snhr-fig}).
This means that the suppression of star formation by continuous SNe II feedback
is stronger than that from instantaneous recycling SNe II.
These results demonstrate that we should not ignore the lifetime of
both SNe II and SNe Ia progenitors,
in studying the dynamical evolution of elliptical galaxies.

\subsection{The Kormendy Relation}
\label{skr}

\begin{figure}
  \leavevmode
  \epsfxsize=80mm  
  \epsfbox{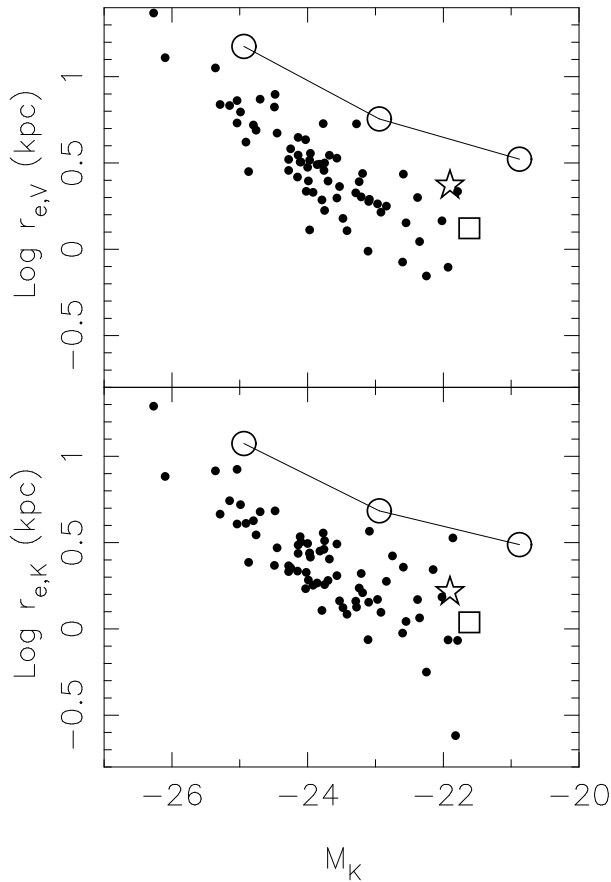}
 \caption{ 
  Comparison of the Kormendy relations for the simulation end-products
 and Coma Cluster galaxies \citep[small dots,][]{mp99} in the $V$
 (upper panel) and $K$ (lower panel) bands.
 The circles connected by solid lines indicate the Kormendy relation 
 for the best model.
 The square (star) denotes the result of model noIa-LM (iIIbIa-HM).
 }
  \label{mre-fig}
\end{figure}

We now examine the physical size of the simulation end-products for all the
models presented above.
Fig.~\ref{mre-fig} shows a comparison of the Kormendy relations
for the simulation end-products and the Coma Cluster galaxies
both in the $V$ and $K$ bands.
We refer to the data of the Coma Cluster galaxies of \citet{mp99}.
In deriving the absolute magnitude and effective
radius in the kpc from the data set in \citet{mp99},
we assume the same distance modulus as mentioned in Section \ref{scmr}.
The best model shows a tendency for higher mass galaxies to have larger 
effective radii, which is qualitatively consistent with the observed
Kormendy relation. However, the effective radii are
systematically larger and the slope is slightly shallower than 
the observational data.
In comparison among the low mass systems, model LM shows 
a significantly larger effective radius  than models noIa-LM and iIIbIa-LM.
Sections \ref{seia} and \ref{seiiia} show that the effect of SNe 
on star formation in the models decreases as: LM, iIIbIa-LM, and
noIa-LM, and the effective radius decreases in the same order.
In other words, stronger feedback causes an expansion of the system.
Due to this expansion effect, the best model fails to explain
the observed Kormendy relation, which is the same problem as 
that found in K01b \citep[see also][]{cc02}.

\subsection{ The [Mg/Fe]--Magnitude Relation }
\label{smgfe}

\begin{figure}
  \leavevmode
  \epsfxsize=80mm  
  \epsfbox{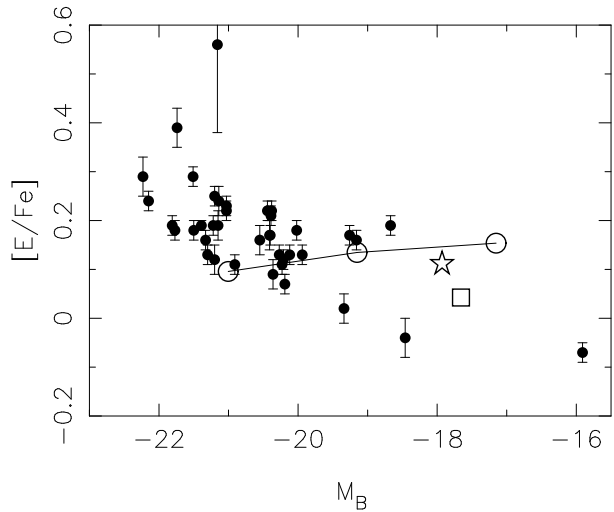}
 \caption{ 
 Comparison of the simulation end-products with observed early-type
 galaxies \citep{tfwg00a} in the [E/Fe]  vs.\ $M_B$ diagram.
 The circles connected by solid lines indicate the the best model.
 The square (star) denotes the result of model noIa-LM (iIIbIa-HM).
 }
\label{mgfe-fig}
\end{figure}

The abundance ratio of [Mg/Fe] of elliptical galaxies has been
well-studied by several groups
\citep{wfg92,ij99,hk00,tfwg00b,tf02}.
Strictly speaking, [Mg/Fe] is not ``observed'', but
is derived by the comparison of the theoretical model with
the observed absorption line features. 
We compare our results with the observational results in
terms of [Mg/Fe] rather than the absorption line features,
because of convenience, although we plan to analyse the absorption 
line features from simulation results and compare them with the observational
data more directly in future work.
The observational results indicate that [Mg/Fe] correlates with
the luminosity, i.e.\ the [Mg/Fe]--magnitude relation.
Because Mg and Fe are mostly produced by SNe II and
SNe Ia, respectively, and because SNe Ia have a longer delay
than SNe II after the formation of stars, this correlation provides a strong
constraint on the star formation history of elliptical galaxies
\citep{mpg98, tgb99, rsmd02, tc02}.
In Fig.~\ref{mgfe-fig}, we compare [E/Fe] for the simulation
end-products of all the models with those derived by \citet{tfwg00a}.
We use the absolute $B$ band magnitudes in \citet{tfwg00b}.
In both the data of \citet{tfwg00a} and the simulation results,
[E/Fe] is measured in the $r_{e,B}/2$ aperture.
Because \citet{tfwg00a,tfwg00b} adopt [E/Fe] instead of
[Mg/Fe] \citep[see also][]{tcb98}, 
we derived [E/Fe] as well as [Mg/Fe]
\footnote{Note that Fig.~12 of K01b compares [Mg/Fe] of the simulation
end-products with [E/Fe] of \citet{tfwg00a}.}. 
The definition of [E/Fe] is as follows:
\begin{equation}
 {\rm [E/Fe]} = ({\rm [Z/H]}-{\rm [Fe/H]})/A,
\label{efe-eq}  
\end{equation}  
where $A=0.929$ for model 4 of \citet{tfwg00a} which we refer.
In model 4 of \citet{tfwg00a}, [E/Fe] indicates the mean enhancement 
of ``E'' group, which contains C, O, Ne, Na, Mg, Si, and S,
with respective to the Fe-peak elements.
The quantity of [E/Fe] should be similar to [Mg/Fe], because
most of the E group elements are nucleosynthetically linked.
Table \ref{ppsp-tab} summarises luminosity weighted [E/Fe] and [Mg/Fe]
for all the model,
and shows that the difference between [E/Fe] and [Mg/Fe] is less than
0.04. Therefore, we consider that it is safe to use the [E/Fe]-magnitude
relation in order to examine the [Mg/Fe]-magnitude relation.
The \citet{tfwg00a} data show a clear tendency
that [E/Fe] increases with the galactic luminosity.
However, the best model is incapable of reproducing this tendency
and has almost constant [E/Fe] regardless of the luminosity.
Models noIa-LM and iIIbIa-LM have similar [E/Fe] to the best model.
As mentioned above, these [E/Fe] abundance ratios
reflect the star formation history
shown in Fig.~\ref{snhr-fig}. In all the models except noIaLM,
the galactic wind causes the cessation of star formation, and
star formation stops before the metal enrichment by SNe Ia progresses.
Since [E/Fe] is determined only by SNe II yields, these models
have similar [E/Fe] to that for model noIa-LM.
 Finally, it is revealed that the best model to reproduce the observed
CMR leads to a short period of star formation and constant [E/Fe],
irrespective of the mass of the system, which is inconsistent with 
the [E/Fe]-magnitude relation derived from observations.
Because the results of [E/Fe] are almost identical to those of [Mg/Fe],
we focus on [Mg/Fe] in the following discussion.

\begin{figure*}
  \leavevmode
  \epsfxsize=160mm  
  \epsfbox{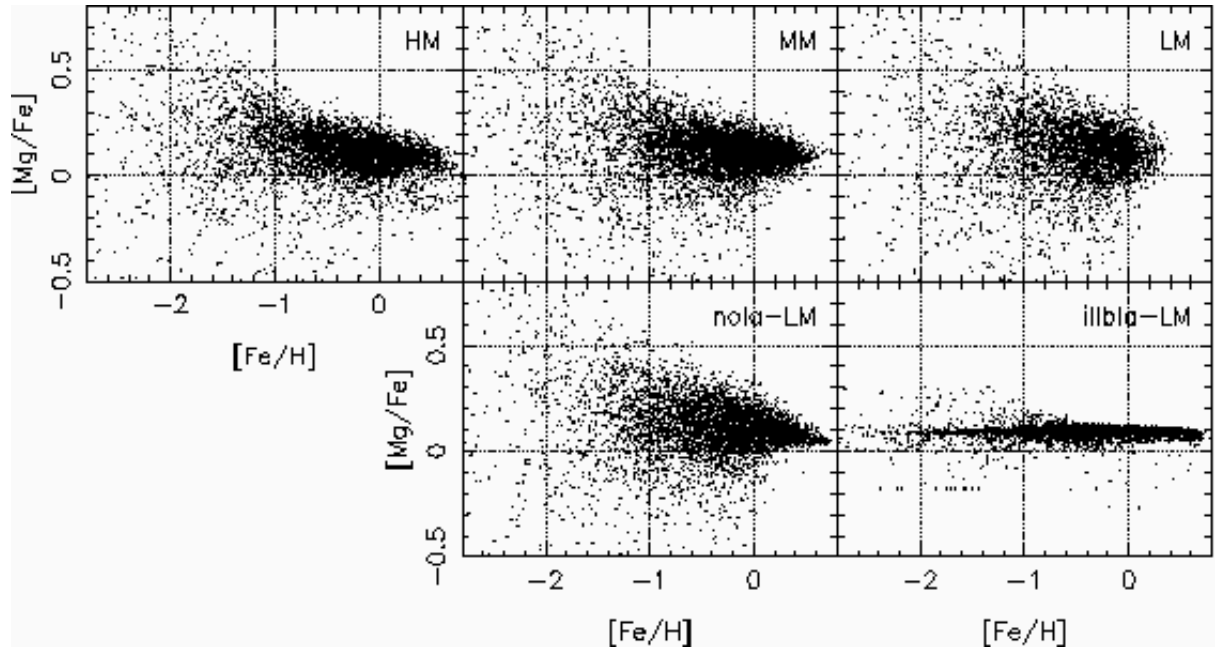}
 \caption{ The [Mg/Fe] against [Fe/H] of the star particles 
 in the final stellar system for all the models. }
  \label{mgfep-fig}
\end{figure*}

A more detailed look shows that the best model has a slight slope 
which is opposite to the [Mg/Fe]-magnitude relation derived from 
observational data, and
model iIIbIa-LM (noIa-LM) has a little lower [Mg/Fe] than model LM
(iIIbIa-LM). These trends also reflect their star formation histories.
Fig.~\ref{mgfep-fig} shows [Mg/Fe] vs [Fe/H] of all star particles
for all the models. In all the models, there is large scatter
in [Mg/Fe] at [Fe/H]$<-1$, and the mean [Mg/Fe] approaches
0.1 at [Fe/H]$>-1$ . 
Model iIIbIa-LM has the least scatter, compared to the other models,
because of the assumption of instantaneous recycling for SNe II.
To interpret Fig.~\ref{mgfep-fig}, 
the mean [Mg/Fe] for the total SNe II yields is useful, which is calculated by
\begin{eqnarray}
 \langle{\rm [Mg/Fe]_{II}}(Z)\rangle & = &
 \log \frac{\int_{8}^{60} M_{\rm Mg,II}(m,Z) \Phi(m) dm}
 {\int_{8}^{60} M_{\rm Fe,II}(m,Z) \Phi(m) dm} \nonumber \\
 & & - \log (Mg/Fe)_{\rm \sun}.
\label{mgfeii-eq}  
\end{eqnarray}  
This value equals [Mg/Fe] of the SNe II yields in model iIIbIa-LM, and
depends on the metallicity, such as 
$\langle{\rm [Mg/Fe]_{II}}(10^{-4} Z_{\rm \sun})\rangle\sim0.3$,
$\langle{\rm [Mg/Fe]_{II}}(10^{-2} Z_{\rm \sun})\rangle\sim0.08$, and
$\langle{\rm [Mg/Fe]_{II}}(Z_{\rm \sun})\rangle\sim0.1$.
Therefore, the [Mg/Fe] versus [Fe/H] distribution in model iIIbIa-LM 
can be easily explained as follows.
Some low metallicity stars with $0.3>$[Mg/Fe]$>0.1$ formed from the
gas enriched by the extremely low metallicity
($Z\sim10^{-4} Z_{\rm \sun}$) stars. The other stars ([Mg/Fe]$\sim0.1$)
are enriched mainly by stars with $Z>10^{-2}$
\footnote{The stars with [Mg/Fe]$=-$0.18 and [Fe/H]$<-1$ are enriched
only by zero metallicity stars, because of 
$\langle{\rm [Mg/Fe]_{II}}(0)\rangle\sim-0.18$.}.
The other models follow the same trend as model iIIbIa-LM,
although they have much larger scatter in [Mg/Fe].
As seen in Fig.\ 3 of \citet{bg97}, [Mg/Fe] of SNe II yields by \citet{ww95}
also depends on the progenitor mass (e.g.\ for solar metallicity
stars, [Mg/Fe]$=$1.2 at 40 M$_{\rm \sun}$ and
[Mg/Fe]$=-$0.59 at 11 M$_{\rm \sun}$). Although the difference
in the life-times of SNe II progenitors is small, this time 
difference causes the inhomogeneous enrichment for the interstellar medium,
and makes the scatter seen in Fig.~\ref{mgfep-fig}.
Some fraction of stars with [Mg/Fe]$<0$ and [Fe/H]$>-1$ 
are enriched by SNe Ia whose yields have [Mg/Fe]$=-1.5$.
Model HM appears to have larger fraction of those stars than
model LM. In addition, model LM stops star formation before enrichment
by SNe II whose yields have [Mg/Fe]$\sim0.1$ becomes dominant.
These two effects lead to the slope of the best model
seen in Fig.~\ref{mgfe-fig}. 

  The reason why [Mg/Fe] of model noIa-LM is smaller than that
of model iIIbIa-LM is a little more complicated -- it is due to
the enrichment from intermediate mass stars.
In theory, with the exception of carbon, nitrogen, and oxygen,
the abundance pattern of the yields for intermediate mass stars
is the same as the abundance pattern
when stars are born, i.e.\ the initial abundance pattern.
However, the abundance pattern for intermediate mass stars 
is set to be the solar
abundance pattern in our code, due to the reason mentioned
in Section \ref{simms}.  As time progresses,
the yields for intermediate mass stars becomes more important,
because more stars with high metallicities finish
the main-sequence phase.
Therefore, model noIa-LM which has the longest duration of star
formation makes a significant amount of stars which suffer from the
yields with the ``artificial'' solar abundance pattern.
These stars are seen in Fig.~\ref{mgfep-fig} as stars with 
high metallicity ([Fe/H]$>0.2$) and slightly lower [Mg/Fe] (around 0.05).
As a result, model noIa-LM has lower mean [Mg/Fe] than models LM
and iIIbIa-LM which have a shorter duration of star formation.
Although inclusion of SNe Ia makes this effect negligible, 
this is caused by the necessarily
simple recipe for the yields for intermediate mass stars, which 
should be improved in the near future.

\subsection{Discussion}
\label{sdisc}

We have applied our new chemo-dynamical evolution code, {\tt GCD+},
to simulations of elliptical galaxy formation, and
confirmed the results of K01b for the CMR,
the Kormendy relation, and the [Mg/Fe]--magnitude relation.
Tables \ref{gpp-tab} and \ref{ppsp-tab} summarise the photometric
properties and stellar population for all the models presented.
The most fundamental difference between our new code 
and the one used in K01b is its taking into account of the lifetime of stars and
the adoption of the new SNe Ia model proposed by KTN00.
We have obtained a more realistic history of SNe Ia rate
than K01b (Fig.~\ref{snhr-fig}). The history of the SNe Ia rate
follows the SNe Ia progenitor model of KTN00.
In the ``best model'' where star formation is stopped
by galactic winds at high redshift, the histories of the SNe Ia rate
have two peaks. These two peaks are caused by SNe Ia from
the MS+WD and RG+WD systems (Section \ref{ssneia}).
Their evolution is similar to that shown in Fig.~5 of KTN00.
After the second peak, this model predicts the continuous explosion of
SNe Ia from the RG+WD system until $z=0$. 
\citet{cet99} provides the observed SNe rates at nearby galaxies.
This SNe II rate for elliptical galaxies gives an upper limit,
and is consistent with the best model where there is no SNe II.
Points with error bars in Fig.~\ref{snhr-fig} show their
observed SNe Ia rates which are transfered from units of SNu 
to units used in Fig.~\ref{snhr-fig} based on the $B$ band luminosity
for each model. In the best model SNe Ia rates are in
good agreement with the observed rates.
This result which comes from a self-consistent chemo-dynamical evolution
provides strong support for the KTN00 model
which is based on pure chemical evolution studies.

Our sophisticated SNe Ia model has clarified that
the feedback effect of SNe Ia plays a crucial role in the evolution
of elliptical galaxies rather than 
that of SNe II. The lifetimes of progenitor stars of SNe II 
are much shorter than those of SNe Ia. As shown in Fig.~\ref{cy-fig},
SNe II occur immediately (between 4 and 40 Myr for stars with Z=0.02)
after star formation events.
On the other hand,
there is a significant delay ($>0.7$ Gyr for stars with Z=0.02)
between star formation and SNe Ia.
Therefore, SNe II occur in dense gaseous environments still actively forming
stars, whereas SNe Ia act on tenuous gas left after star formation.
This difference makes SNe Ia the major trigger of galactic winds.
Comparison between models noIa-LM and LM shows this clearly in Section
\ref{seia}. Also model iIIbIa-LM demonstrates that the simplified model
for SNe II and SNe Ia underestimates significantly 
the SNe effect on the dynamical evolution of elliptical galaxies.

 The difference in SNe II and Ia model between K01b and this paper
is one of the reasons why our new model 
requires only $10^{50}$ ergs and $f_v=0.002$ as an SN feedback energy
to reproduce the observed CMR, 
whereas K01b requires $4\times10^{51}$ ergs and $f_v=0.9$.
This difference is caused by the combination of
some physical processes, such as the IMF, SNe II and SNe Ia feedback,
and the cooling function, which are updated in the new code.
Comparing the results, our new code requires a more reasonable
amount of the SN feedback energy than that in K01b, 
to reproduce the observed CMR.
However, we might still ignore some physical processes which 
contribute to the CMR, and be able to further reduce the feedback energy.
For example, the UV background radiation
is a possible candidate as such physical process, as suggested by K01b.
A recent semi-analytical study demonstrates that the UV background
radiation has a similar effect to the SNe feedback 
on the CMR \citep{ng01}.

 We find that mass dependences of both the galactic wind and 
the gas infall rate are responsible for the slope of the CMR.
The conventional galactic wind scenario considers only the former mechanism.
Our numerical simulation shows that 
the higher mass system has a higher virial temperature and
lower efficiency of radiative cooling.
Therefore, the gas infall rate decreases with increasing mass 
of the system.
This lower gas infall rate leads to the increase in the fraction of the high
metallicity stars (Fig \ref{meta-fig}). As a result,
the mean metallicity of the higher mass system
becomes higher than that of the lower mass system.
This is a novel mechanism contributing to the CMR.

 Also in the [Mg/Fe]--magnitude relation, we reach the same conclusion
as K01b qualitatively. However, the value of [Mg/Fe] is
systematically different from that in K01b. This is due to the difference in
the stellar yields of SNe II. We use the yields
calculated by \citet{ww95}, whereas K01b adopts those by \citet{nht97}.
Although \citet{ww95} and \citet{nht97} are two representative
computations for SNe II nucleosynthesis,
it is known that the amount of iron yield derived in \citet{ww95} 
is significantly larger than that in \citet{nht97}, 
and therefore the yields of \cite{ww95}
lead to smaller [Mg/Fe] \citep{tww95, bg97, glm97,tgb98}.
The theoretical stellar yields still have a large uncertainty, and
to discuss the zero-points of [Mg/Fe] for simulation end-products
we need a more reliable yield model.
Nevertheless, we can discuss the observed slope 
of the [Mg/Fe]--magnitude relation, and our model so far cannot explain 
this.

To explain the observed slope, it is a plausible scenario
that low mass systems have a longer duration of star formation
and are enriched  by SNe Ia more than high mass systems
 \citep{fwg92,fm94,mpg98}.
K01b suggests that the UV background radiation is a possible candidate
to realise the scenario,
because it suppresses cooling and star formation more 
strongly in lower mass systems \citep{ge92},
and is expected to prolong the duration of star formation.
Since we have developed a sophisticated chemo-dynamical evolution code,
it is important to introduce the UV background radiation
into {\tt GCD+}, and examine the effect of the UV background radiation
on galaxy evolution.
In a forthcoming paper, we hope to throw a light on this issue.

\section{Conclusion}
\label{sconc}

 We have developed a new chemo-dynamical evolution code
called {\tt GCD+}.
The code includes both SNe II and SNe Ia modeling taking into
account the lifetime of progenitor stars, and
the chemical enrichment from intermediate mass stars.
In this paper we have applied the code to simulations of elliptical galaxy 
formation described as the evolution of an isolated seed galaxy.
We have shown that {\tt GCD+} is 
a useful and unique tool which enables us 
to compare simulation results with
the observational data directly and
quantitatively, and gives us new insights into the
effects of dynamics on the chemical evolution of galaxies.
Encouraged by the success of this study, we plan to apply {\tt GCD+} to 
high-resolution cosmological simulations in the near future.
In the real universe, 
even isolated galaxies at $z=0$ are expected to experience 
interactions with the another galaxies in the past. In fact, \citet{ng98}
suggests that interactions between galaxies are 
a dominant mechanism to transport metals from the galaxies to
intergalactic medium rather than SNe feedback. 
Therefore, using simulations covering a cosmological scale,
it will be valuable to examine 
what physical process, or combination of processes, 
induces the tight observed CMR. 
In addition, cosmological simulations using {\tt GCD+} can provide detailed
metallicity distributions not only in the stellar component
of individual galaxies, but also in the intergalactic medium,
such as Lyman-$\alpha$ clouds and the hot gas in clusters of galaxies.
As mentioned in Section \ref{sfd}, the different chemical elements
have progenitors with different masses,
and different mass stars have different lifetimes.
It is highly probable that the abundance ratios are fairly sensitive to
the metal transport mechanism and its epoch.
New observational facilities will offer a wealth of 
information about the distribution of metallicity, as well as the
abundance ratio of each element, such as [N/O], [Mg/Fe] and [Si/Fe],
with unprecedented accuracy.
Studies comparing new observations with
our future cosmological simulation results 
will show clearly which metal transport mechanism is dominant,
and thus contribute to understanding the formation and evolution of
galaxies as well as clusters of galaxies from a new perspective.

\section*{Acknowledgments}

We thank Michael A. Beasley for his helpful advice during
the completion of this manuscript, and the referee for their thorough 
reading of the manuscript.
We are grateful to Nobuo Arimoto and Tadayuki Kodama
for kindly providing the tables of their SSPs data.
We are also grateful to Edmund Bertschinger for generously providing
the COSMICS program.
We acknowledge the Yukawa Institute Computer Facility,
the Astronomical Data Analysis Center of the National Astronomical
Observatory, Japan, the Australian and Victorian Partnership for Advanced
Computing, where the numerical computations for this paper were performed.
This work is supported in part by
the Australian Research Council through the Large Research Grant 
Program (A0010517), Japan Society for the Promotion of Science
through the Grants-in-Aid for Scientific Research (No.\ 14540221),
and Swinburne University through the Research Development Grants Scheme.

\label{lastpage}

\end{document}